\documentstyle[spie]{article} 
\input{psfig}   
\title{Cancellation of GLONASS signals from Radio Astronomy Data} 

\author{Steven W. Ellingson\supit{a}, John D. Bunton\supit{b}, Jon
F. Bell\supit{c} 
\skiplinehalf 
\supit{a}OSU ElectroScience Laboratory 1320 Kinnear Road, Columbus OH 43212
USA \\
\supit{b}CSIRO Telecommunications and Industrial Physics, PO Box 76 Epping
NSW 1710, AUSTRALIA\\
\supit{c}CSIRO Australia Telescope National Facility, PO Box 76 Epping NS
W 1710, AUSTRALIA \\
}

\authorinfo{Further author information: (Send correspondence to S.W.E. or
J.F.B.)\\S.W.E. : E-mail: ellingson.1@osu.edu~~~~~J.F.B. : E-mail:
jbell@atnf.csiro.au~~~~~J.D.B. : E-mail: john.bunton@tip.csiro.au}

\begin{document} 
\maketitle 

\begin{abstract}
Astronomers use the 1612 MHz OH spectral line emission as a unique window on
properties of evolved stars, galactic dynamics, and putative proto-planetary
disk systems around young stars. In recent years, experiments using this OH
line have become more difficult because radio telescopes are very sensitive
to transmissions from the GLONASS satellite system. The weak astronomical
signals are often undetectable in the presence of these unwanted human
generated signals.  In this paper we demonstrate that GLONASS narrow band
signals may be removed using digital signal processing in a manner that is
robust and non-toxic to the weak astronomy signals, without using a reference
antenna. We present results using real astronomy data and outline the steps
required to implement useful systems on radio telescopes.
\end{abstract}


\keywords{GLONASS, interference, cancellation, radio astronomy}

\section{INTRODUCTION}
\label{sec:intro}  

Many papers in the astronomical literature cite problems with interference
from the Russian {\em Global'naya Navigatsionnaya Sputnikovaya Sistema}
(GLONASS) system of navigational satellites when trying to observe 1612 MHz
OH spectral line emission. Galt (1991) \cite{gal91} and Combrinck et
al. (1994) \cite{cwg94} both present data demonstrating the damaging effect
of GLONASS signals on astronomy data.  Some reports have stated that up to
50\% of all observations have had to be discarded \cite{gal91}.  The
scientific merits of OH spectral line observations are discussed in detail
elsewhere \cite{coh89,hh85}; however, there is no question that this is
extremely valuable spectrum whose continued use is essential to radio
astronomy.

One possible solution to the problem is regulation; this is being addressed
within international organisations such as the ITU and URSI.  However,
regulation cannot be expected to recover the spectrum into which the GLONASS
system already transmits. The solution most often employed by radio
astronomers in dealing with unwanted signals is to put their telescopes in
remote locations. However, when dealing with signals that emanate from
Earth-orbiting satellites, that method obviously fails. The next most
obvious solution is not to observe when interfering signals are present, or
simply throw away affected data \cite{gal91}. Some ``GLONASS aware'' tools
have been developed that allow dynamic scheduling observations in order to
minimise interference \cite{cwg94}. However, the strategy of avoidance
results in the loss of valuable telescope time, which often amounts
thousands of dollars per day, a better solution is desired.

Here we present a direct, technical solution to the problem.  We have
developed and demonstrated a parametric signal processing algorithm which
identifies GLONASS signals present in the pre-detection, complex baseband
telescope output, and removes them.  This algorithm results in a high degree
of suppression with negligible distortion of radio astronomical signals.  We
believe this approach can be applied to interference from the U.S. Global
Positioning System (GPS) and possibly other sources as well.  This technique
is presented in Section~3 of this paper.  First (Section~2), we describe the
properties of GLONASS that are relevant to the operation of the canceller.
In Sections~4--5, we present experimental results demonstrating the
effectiveness of this approach.  Section~4 describes the procedure used to
collect GLONASS-corrupted data; whereas Section~5 shows the results before
and after application of the canceller. In section~6 we consider how this
approach may be implemented on existing telescope systems.

\section{Properties of GLONASS signals}

GLONASS satellites transmit at frequencies between 1602--1616~MHz and have
shared primary user status with radio astronomy for the 1610.6--1613.8~MHz
band\cite{cwg94}. There are 24 carriers spread over the 14~MHz band at
intervals of 0.5625 MHz. The carrier is modulated by a pair of noise like,
equal power, pseudo noise (PN) codes of 0.511 and 5.11 MHz. Figure~1. of
Combrinck et al. (1994)\cite{cwg94} shows time averaged spectra of these
signals. The unfiltered sinc$^{2} $ side lobes of these signals have
relative power levels as high as $-25$~dB extending out to 20 MHz either
side of the main carrier in some cases\cite{gal91}. GLONASS satellites
launched more recently do have some band-limiting filters.

GLONASS, despite its wide band spectrum, actually has a very simple
structure\cite{GLONASS_ICD}.  Consider the narrow band (0.511~MHz) GLONASS
modulation.  This signal is simply a sinusoidal carrier which experiences a
phase shift of $0^{\circ}$ or $180^{\circ}$ every
$(0.511~\mbox{MHz})^{-1}$. Each phase shift represents a modulation symbol,
or {\em chip}.  Each group of 511 chips represents a PN code, which is
public knowledge, never changes, and is the same for every GLONASS
satellite. GLONASS data bits are represented by changing the sign of a block
of 10 PN codes, with 10~ms period.  Parameters of the signal which are
unknown when received are (1) the Doppler shift due to satellite motion, (2)
the {\em code phase}, that is, the relative position within the 1~ms PN code
period, and (3) the carrier phase, which rotates because the satellite is
moving and the transmitter's LO is not perfectly stable.  However the
carrier phase, the current value of the data bit, and the complex gain due
to the antenna pattern can all be combined into a single unknown complex
magnitude parameter.  Thus three parameters are sufficient to describe the
GLONASS signal with high accuracy.

Finally, we note that the modulation used by the course/acquisition (C/A)
mode of the U.S. Global Positioning System (GPS) is very similar to
modulation used in the GLONASS 0.511~MHz transmission.  The main differences
are longer code (1023~chips) and higher chip rate (1.023~MHz); also, all GPS
satellites transmit on the same centre frequency, but with different (but
known) PN codes.  Thus, techniques which are effective against 0.511~MHz
GLONASS modulation may be effective against GPS C/A transmissions as well.

\section{CANCELLATION ALGORITHM}

\subsection{Theory}

\begin{figure}[htbp]
\begin{center}
\begin{tabular}{c}
\psfig{figure=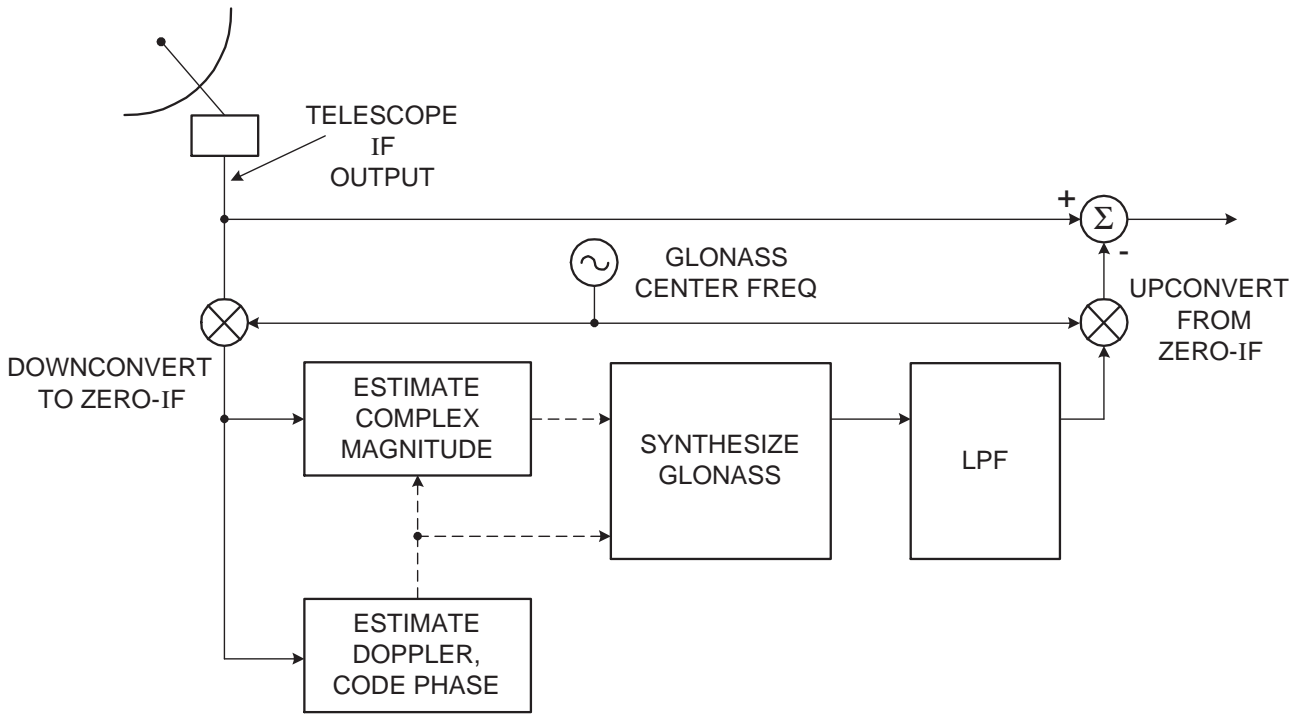,height=9.5cm} 
\end{tabular}
\end{center}
\caption{Parametric cancelling technique. The ESTIMATE DOPPLER, CODE PHASE
block uses the down converted telescope signal to generates an estimate of
the time aligned PN code and the offset in frequency between the GLONASS
CENTER FREQ oscillator an the down converted GLONASS signal.  These two
estimates and the down converted signals are used by the ESTIMATE COMPLEX
MAGNITUDE block to calculate a magnitude and phase correction.  These
estimates of time aligned PN code, frequency, magnitude and phase are then
used to synthesise a zero-IF GLONASS signal.}
\label{fPSM}      

\end{figure}

Our technique for suppressing GLONASS signals in radio astronomy data is
based on {\em parametric signal modelling}.  Recall that the GLONASS signal
can be described using a model consisting of just three parameters: Doppler,
code phase, and complex magnitude.  Given a block of data containing a
GLONASS signal, one can then estimate the parameters.  Given the parameters,
it is possible to synthesise a noise-free copy of the GLONASS signal.  This
copy is then subtracted from the telescope output to achieve the
suppression. This procedure is illustrated in Figure~\ref{fPSM}.

The parametric solution proceeds on two time scales.  Doppler frequency and
code phase are difficult to estimate, but change slowly.  Complex magnitude,
on the other hand, is simple to estimate, and changes quickly.  Our approach
is to first {\em acquire} the GLONASS signal.  This involves a joint search
over the possible Doppler frequencies and the code phases.  For each
Doppler/code phase pair, a complex baseband (zero-IF) version of the signal
is cross-correlated with the PN sequence.  The correct Doppler/code phase
pair is the one which maximises the magnitude of the cross-correlation.
Although tedious, this is a simple procedure, and is essentially the same
acquisition procedure used by hand held GPS receivers.  Once acquired, the
Doppler and code phase can be tracked simply by sensing the drift in the
correlation peak and adjusting the Doppler and code phase parameters
accordingly.  It appears that the Doppler and code phase estimates can be
frozen for at least 0.1s between updates without any significant effect on
the results.

Once the signal is acquired, we estimate the complex magnitude by
cross-correlating the time- and frequency-aligned PN code with the complex
zero-IF representation of the GLONASS signal.  The magnitude and phase of
the cross-correlation then represents the desired complex gain. The complex
gain is expected to change quickly, so this procedure must be updated often.
In the example presented below, the complex gain update rate is 128~$\mu s$,
using 1024 samples at 8~Msamples per second (conversion to a complex
baseband signal has halved the sample rate).

Given the Doppler frequency, code phase, and complex magnitude, one can then
synthesise a noise-free estimate of the GLONASS signal. However, it has been
found by experience that better cancellation is achieved by low-pass
filtering the zero-IF version of the synthesised GLONASS signal before
subtraction from the telescope output.  This models the real-world low-pass
effect which smoothes discontinuities in band limited signals. This also has
the desirable effect of suppressing the high-order side lobes of the
synthesised signal, which may not be accurately represented by the proposed
signal model.  A suitable filter was found to be a 32-tap finite impulse
response (FIR) filter based on the Hamming window, with cutoff frequency
equal to $0.05F_{S}$ at $F_{S}=8$~MSPS.  Such a filter can be obtained
using the MATLAB command {\tt fir1(32,0.1)}.

\subsection{Implementation}

The results presented below were obtained using non-real-time
post-processing software, written in MATLAB. On a 400MHz pentium the
processing presently runs at 1000 times real time. The MATLAB source code is
freely available from the authors. Any practical system would, of course,
require real time implementation. The maturity of GPS technology means that
the techniques and hardware for the acquisition of GLONASS and GPS signal
parameters are well developed.  The design of the signal modulators in the
satellites is also known.  With the knowledge of these two areas a practical
real time implementation is within reach and is discussed in Section~6.


\section{DATA COLLECTION} 

The astronomy data used in testing these algorithms is a single linear
polarisation, 4-bit data stream from each antenna of the 6x22m antenna,
CSIRO Australia Telescope Compact Array at Narrabri in Australia. The data
was 4-bit sampled at 16MHz and recorded on an S2 recorder\cite{s2ref}. The
resulting 8MHz bandpass centred on 1610 MHz was wide enough to include
signals from GLONASS-69 at $\sim$1609~MHz, an OH maser source (IRAS 1731-33)
at $\sim$1612~MHz and some flat spectrum. The data were then extracted using
the S2TCI system\cite{s2ref} and demultiplexed. More details on this dataset
and others that are freely available for conducting these kind of
experiments are in reports by Smegal et al.\cite{sme99} and Bell et
al.\cite{bel99}. The algorithm works on a single polarisation data stream
from one antenna only. However, data from a second antenna were also used in
cross correlations as a test of how well the GLONASS signals were removed.

\section{RESULTS TO DATE}

\begin{figure}[htbp]
\begin{center}
\begin{tabular}{c}
\psfig{figure=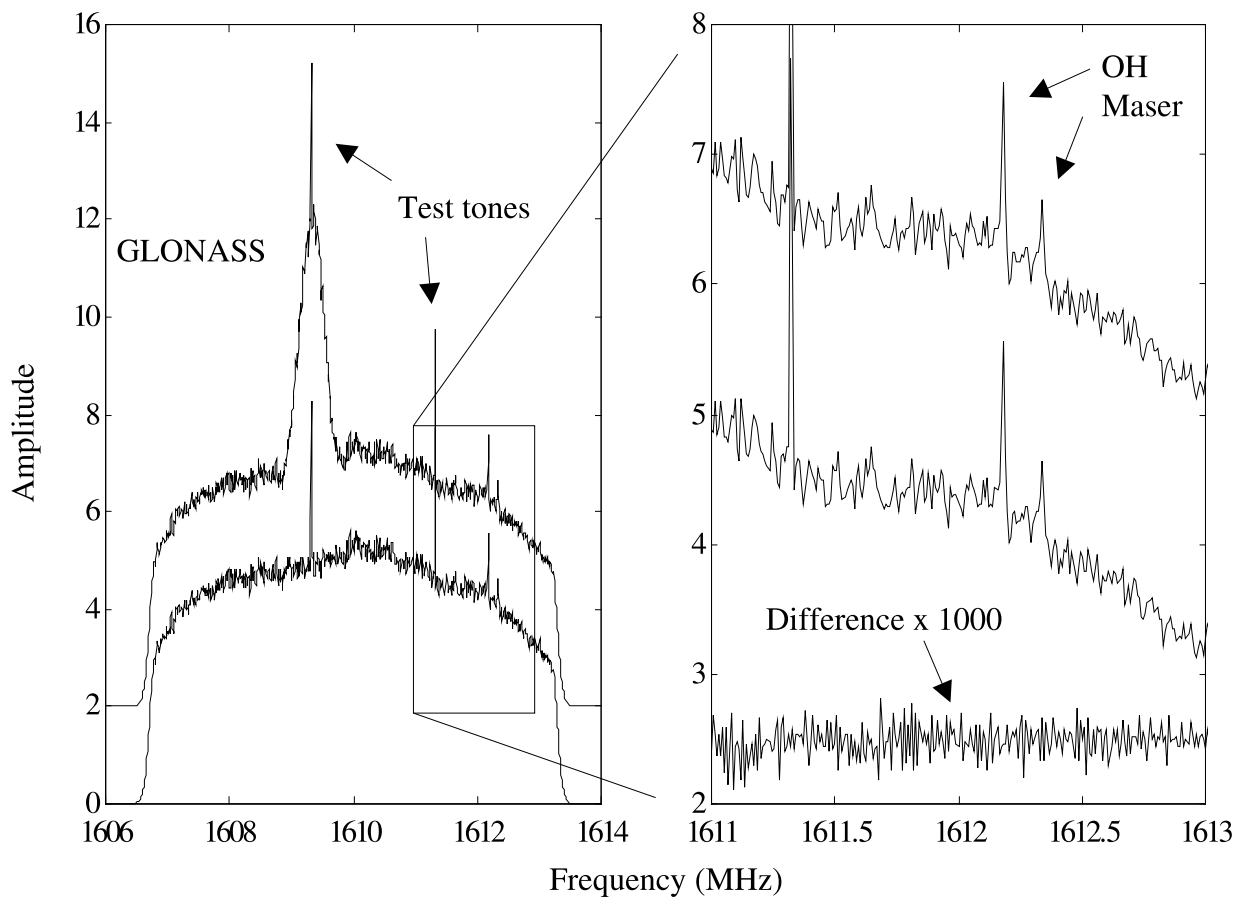,height=13cm,angle=-90} 
\end{tabular}
\end{center}
\caption{Left plot: top curve: raw data, OH maser source at 1612 MHz, test
tones inserted (in software) at 1609.3 and 1611.3 MHz. Left plot: lower
curve: data with GLONASS removed. Right plot: blow up of region around OH
source. The bottom curve shows the difference of the pre and post
cancellation spectra multiplied by 1000. This indicates that this region of
the spectrum is not changed by more that a few parts in 1000. (curves offset
for clarity)}
\label{fig:skeptics}      
\end{figure}

The results so far are encouraging, with GLONASS narrow band signals being
effectively removed in a way that is non toxic to astronomy signals. Figure
\ref{fig:skeptics} (left plot, top curve) shows a spectrum of the raw data,
with test tones added in software. The bottom curve shows the same 0.1s
($1.6 \times 10^6$ samples) of data with cancellation technique applied,
with no apparent GLONASS signals left. There is an OH maser source at
$\sim$1612~MHz. The top two curves in the right plot of Figure
\ref{fig:skeptics} show a blow up of this region. The bottom curve in the
same plot shows the difference multiplied by a factor of 1000, indicating
that no damage has been done in the spectral region of the OH source.

\begin{figure}[htbp]
\begin{center}
\begin{tabular}{c}
\psfig{figure=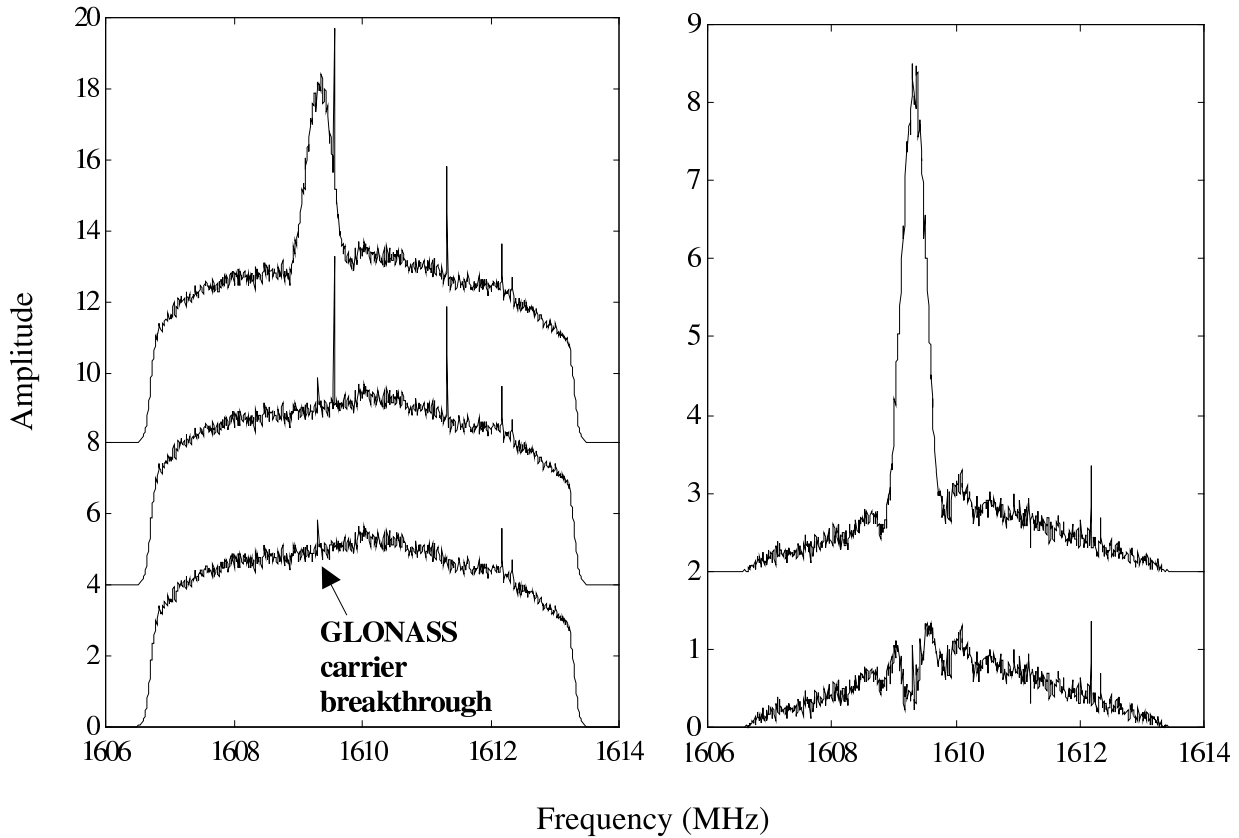,height=13cm,angle=-90} 
\end{tabular}
\end{center}
\caption{Left plot: Top: Offset test tone added before cancellation. Middle:
Carrier signal break through apparent after GLONASS signal has been
cancelled. Bottom: Offset test tone subtracted after cancellation shows the
test tone is unaffected.  Right plot: top curve: Cross correlation of raw
data from 2 antennas. Bottom curve: Cross correlation of data with GLONASS
removed and raw data from another antenna. (Curves offset for clarity).}
\label{fig:tt_xcorr}         
\end{figure}

In order to examine the toxicity of this algorithm in the same frequency
range as the GLONASS signal we added a test tone to the data before the
cancellation was applied and examined how it was affected by the
processing. The left plot of Figure \ref{fig:tt_xcorr} shows the result of
subtracting the test tone again, after the cancellation. There is no
evidence that the test tone has been affected by the cancellation. However,
there is a small spike left right under the middle of the GLONASS signal,
that is unrelated to the test tones. This seems to be a result of some break
through, or leakage of the GLONASS carrier signal from the GLONASS imbalance
in the GLONASS phase modulator. It should be possible to model and remove
this as well, but we have not addressed that yet.

A more sensitive test of the supression is to cross correlate with signals
from another antenna. The Right plot of Figure \ref{fig:tt_xcorr} show some
cross correlations. The top curve is the cross correlation of raw data from
two antennas. The bottom curve is the cross correlation of raw data from one
antenna and data with GLONASS cancelled from data from another
antenna. There are some extra ripples here that are not apparent in the
other spectra, suggesting that there are some inaccuracies in the algorithm
that need further investigation. The majority of the signal seen in this
cross correlation is probably due to the GLONASS wide band signal, which we
have not tried to suppress yet.

\begin{figure}[tbp]
\begin{center}
\begin{tabular}{c}
\psfig{figure=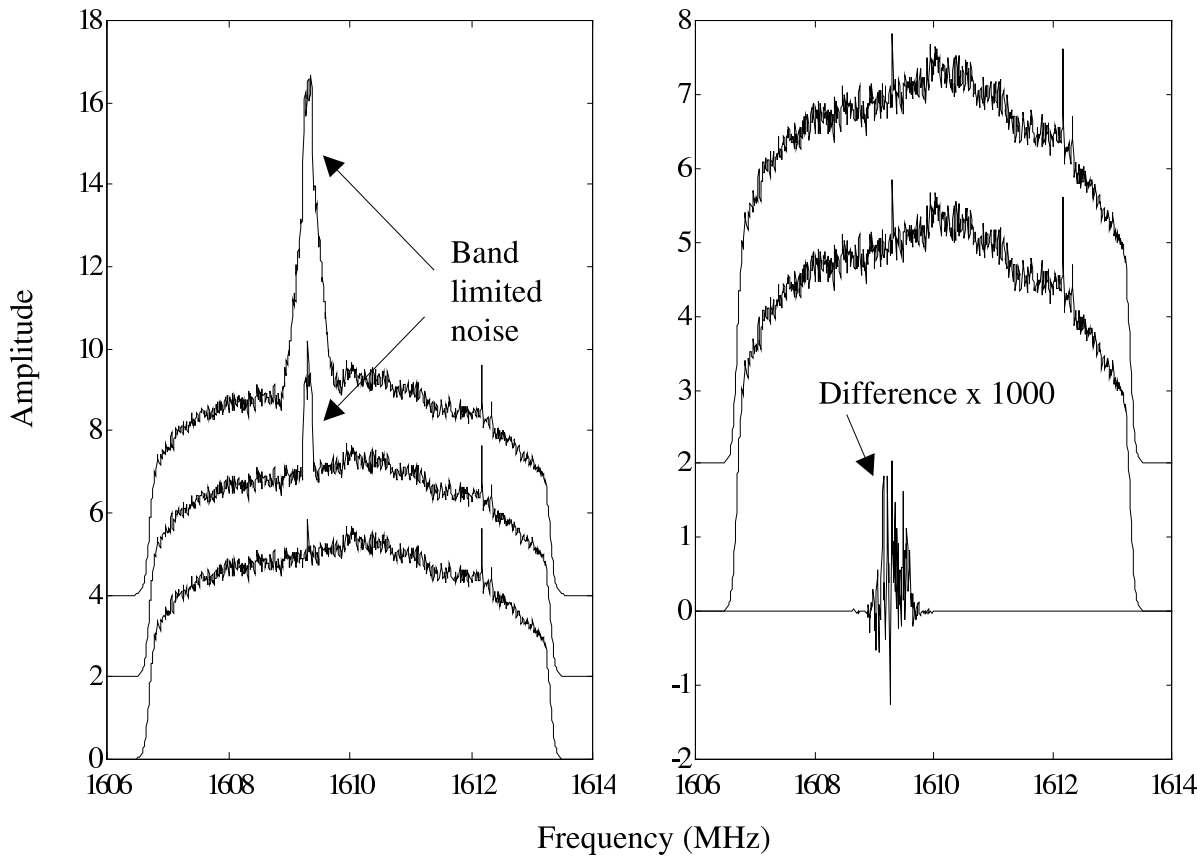,height=13cm,angle=-90} 
\end{tabular}
\end{center}
\caption{Left plot: Left plot: Top: Band limited noise added before
cancellation. Middle: GLONASS signal has been cancelled. Bottom: Band
limited noise subtracted after cancellation shows the band limited noise is
unaffected by the cancellation (the carrier break through is still
present).  Right plot: Top: Result of cancellation with no test tone or
noise added. Middle: Result of cancellation with bland limited noise added
before and subtracted after cancellation. Bottom: Difference of top 2 curves
multiplied by 1000 shows that the band limited noise was not affected by more
than 1 part in 1000. (Curves offset vertically for clarity in several
cases).}
\label{fig:blnoise}         
\end{figure}

The addition and subtraction of test tones give us some indication of how
toxic the algorithm is to astronomy signals. However, astronomy signals are
not coherent sine waves, but are more like band limited noise in the case of
spectral lines and wide band noise in the case of continuum sources. We
replaced the test tones with some synthetic band limited noise, added it
before the cancellation and subtracted it again after. As shown in Figure
\ref{fig:blnoise} the band limited noise is not affected by the cancellation
algorithm, down to a part in 1000, in other words, we have achieved 30~dBs
of dynamic range in the supression. Processing with and without the presence
of an astronomy sources make very little difference to the effectiveness of
the cancellation.

\section{LIMITATIONS AND FUTURE IMPROVEMENTS}

The carrier break through described earlier needs to be modelled and
cancelled. While we have not investigated this carefully yet, we believe
this should be possible. We do not know the cause of the ripples in the
cross correlation spectra and some more investigation there may lead to an
improvement in the algorithm. At present we have modelled the band limiting
filters of the GLONASS transmission system by a low-pass filter when the
signal is centered on zero-IF. This is limiting the accuracy with which we
estimate the GLONASS spectrum and therefore possibly limiting the
cancellation. We aim to either obtain details of the band limiting filters
used on the GLONASS transmission system, or use the existing data to
estimate them.

Note that both GPS and GLONASS have in-band secondary channels that are
spread 10 times wider, and therefore have power spectral density about 10
times weaker. This secondary channel for GPS and GLONASS is much harder to
deal with, because the PN codes are more complex and may not be completely
known. We do need to find a way to mitigate against these signals, because
they do still cause substantial problems for radio astronomy.

In addition to GLONASS and GPS, many other satellite systems have
well-specified modulation and coding schemes.  This opens up the
possibility of removing these other classes of signals using digital signal
processing. The techniques described here will be useful for some others,
but not all.  At a recent meeting\cite{efwhite} a wide range of interference
mitigation techniques applicable to many different undesired signals were
presented and discussed.

The usefulness of the method presented in this paper must ultimately be
shown by conducting astronomy in the presence of GLONASS or GPS.  Ideally
the astronomy results with and without interference present would be
indistinguishable.  The use of recorded data and post processing is useful
for demonstrating the method.  But with processing currently running at one
1/1000 real time the amount of astronomy that can be demonstrated is
very limited.  It is therefore desirable to explore the use of dedicated
hardware to achieve real time processing of the signal.

The most comprehensive approach is to build a complete receiver that
incorporates the interference cancellation method described in this paper.
This is the approach that must be taken in designing new interference
resistant receivers but currently the best option is to build an 'add on' to
existing receivers.  The problem with this method is that the quantiser in
current receivers is designed for adequate performance when processing noise
like signals. Typically a 1 or 2 bit quantiser is used.  Techniques like
adaptive noise cancellation will cause the number of bits needed to
represent the signal to grow when the interference is suppressed.  For
example, consider an interferer whose peak amplitude is equivalent to one
eighth of the least significant bit in the quantiser and assume that 2 bits
can accurately represent the interference.  To generate an interference free
estimate of the signal it is necessary to subtract the estimate of the
interferer from the signal.  In doing this, the number of bits needed to
represent the signal grows by 4 lower order bits.  The signal now has too
many bits to be processed by receivers normally used for radioastronomy.
The solution to this problem is to form auto and cross correlations of the
measured signal and the estimate of the interference.  If the estimate e(t)
and the interference i(t) are related to each other by a linear transfer
function then in frequency domain this can be written as I(f) = H(f)E(f)
where H(f) is the frequency domain transfer function.

In this context the astronomy signal can be considered to noise.  The system
can be redrawn as

\begin{center}
\begin{tabular}{c}
\psfig{figure=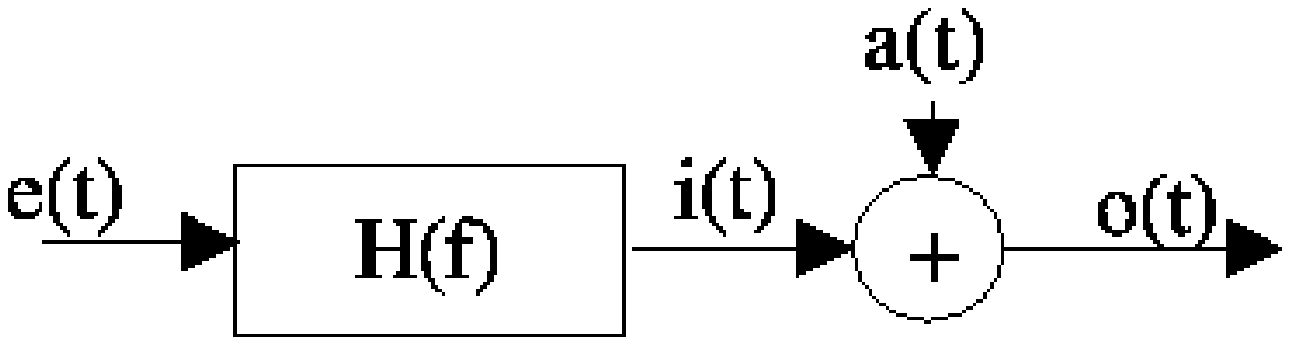,height=2.5cm,angle=-90} 
\end{tabular}
\end{center}

where o(t) is the wanted astronomy signal plus interference. The cross-spectrum method can now be used to estimate the transfer function
H(f) .  If the astronomy signal a(t) is uncorrelated with i(t) then
H(f)\cite{bp71,pap89} equal to the ratio of the cross-spectrum $S_{oe}(f)$
between o(t) and e(t) and the power spectrum $S_{ee}(f)$ of e(t).

$H(f) = S_{oe}(f)/ S_{ee}(f)$ 

The power spectrum $S_{aa}(f)$ of a(t) is now equal to: 

$S_{aa}(f) = S_{oo}(f)-|H(f)|^{2}.S_{ee}(f)$

Thus measurement of the power spectrum of a(t) and e(t) plus the cross
spectrum of the two gives enough information to derive the power spectrum of
the uncorrupted astronomical signal.  These spectra can be obtained from the
auto and cross correlations of the two signal.  Correlators used for
astronomy normally process at most 2-bit data.  Straight 2-bit sampling may
not be sufficient to accurately represent the estimated GLONASS/GPS
signal. The addition of dither and the use of noise shaped oversampling
should solve this problem.  The hardware itself will internally generate a
multi-bit accurate representation of the interferer.

The hardware used to synthesise the baseband GLONASS or GPS is comparatively
simple and easy to emulate in an FPGA.  The main difficulty is the
estimation of and tracking of signal phase and amplitude.  This task is best
left to software.  Data needed to perform this task are correlations between
the input and the current estimate of the interferer. If a reasonable
initial estimate of the interferer has been found then very few correlations
are needed to maintain tracking of carrier phase, carrier Doppler and code
phase.  The hardware could also be used to generate these correlations. The
operations performed by the hardware are:

\begin{enumerate}
\item Generate the carrier digitally with adjustable carrier phase and Doppler.
\item Generate the GLONASS/GPS modulation with adjustable code phase
\item Modulate the carrier with the GLONASS code. This gives an unscaled
'noise free' estimate of the interference.
\item Generate the zero delay and $\pm1/2$ chip correlations between the
input signal and the 'noise free' estimate.  
\item Scale the magnitude of the 'noise free' estimate to match the
interference.
\item Optionally delay data and estimate. This allows the magnitude and phase corrections to be applied to the estimate used in generating the corrections. This extra item is needed to fully emulate the current software.  In practice it may be unnecessary.
\end{enumerate}

This hardware removes most of the intensive tasks from software and
leaves the software to monitor the correlations.  From this monitoring the
software then needs to generate updates for the carrier phase, carrier
Doppler, code phase and amplitude.  With these updates the output of the
hardware is a 'noise free' estimate of the interference. 

\subsection*{Acknowledgements}
The authors gratefully acknowledge discussions with and help in
obtaining data from Ron Ekers, Rick Smegal, Peter Hall, Bob Sault,
Matthew Bailes, Willem van Stratten, Frank Briggs, Mike Kesteven, Warwick
Wilson, Dick Ferris

\end{document}